\documentclass[aps,prl,amsmath,amssymb,onecolumn,superscriptaddress,showpacs]{revtex4}
\usepackage{amsmath,graphicx,float,epsfig,float,tabularx,multirow}
\usepackage{epstopdf,color,bbm,times}
\usepackage[english]{babel}

\begin{document}
\title{Coherence properties of high-gain twin beams generated in pump-depletion regime}
\author{A.~Allevi}
\email{alessia.allevi@uninsubria.it} 
\affiliation{Dipartimento di Scienza e Alta Tecnologia, Universit\`a degli Studi dell'Insubria and CNISM UdR Como, Via Valleggio 11, I-22100 Como, Italy,}
\author{O.~Jedrkiewicz}
\affiliation{Istituto di Fotonica e Nanotecnologie, Consiglio Nazionale\\ delle Ricerche and CNISM UdR Como, Via Valleggio 11, I-22100 Como, Italy,}
\author{ E.~Brambilla}
\affiliation{Dipartimento di Scienza e Alta Tecnologia, Universit\`a degli Studi dell'Insubria and CNISM UdR Como, Via Valleggio 11, I-22100 Como, Italy,}
\author{A.~Gatti}
\affiliation{Istituto di Fotonica e Nanotecnologie, Consiglio Nazionale\\ delle Ricerche and CNISM UdR Como, Via Valleggio 11, I-22100 Como, Italy,}
\author{J.~Pe\v{r}ina,~Jr.}
\affiliation{ RCPTM, Joint Laboratory of Optics of Palack\'{y} University and Institute of Physics of Academy of Sciences of the Czech
Republic, Faculty of Science, Palack\'{y} University, 17.listopadu 12, 77146 Olomouc, Czech Republic}
\author{O.~Haderka}
\affiliation{ RCPTM, Joint Laboratory of Optics of Palack\'{y} University and Institute of Physics of Academy of Sciences of the Czech
Republic, Faculty of Science, Palack\'{y} University, 17.listopadu 12, 77146 Olomouc, Czech Republic}
\author{M.~Bondani}
\affiliation{Istituto di Fotonica e Nanotecnologie, Consiglio Nazionale\\ delle Ricerche and CNISM UdR Como, Via Valleggio 11, I-22100 Como, Italy,}

\begin{abstract}
Twin-beam coherence properties are analyzed both in
the spatial and spectral domains at high-gain regime including pump
depletion. The increase of the size of intensity auto- and
cross-correlation areas at increasing pump power is replaced
by a decrease in the pump depletion regime.
This effect is interpreted as a progressive loss in the mode selection
occurring at high-gain amplification. The experimental
determination of the number of spatio-spectral modes from $
g^{(2)} $-function measurements confirms this explanation.
\end{abstract}

\pacs{42.65.Lm Parametric down conversion and
production of entangled photons, 42.65.-k Nonlinear optics, 42.50.Ar Photon statistics and
coherence theory, 85.60.Gz Photodetectors}

\maketitle

The process of parametric down-conversion (PDC) in bulk nonlinear
crystals generates twin-beam states of light that are naturally
multi-mode both in spectrum and space \cite{Allevi12}. Many works,
performed in the single-photon regime, have highlighted the
correlations and coherence properties of photon pairs either in
spectrum or space
\cite{Hamar2010,Blanchet08,Moreau12,Dayan04_05,O'Donnel09,Sensarn10,Haderka2005a,PerinaJr2012}.
In the past ten years, also high-gain PDC, leading to a large
number of photons per mode, has been the object of several studies
for its interesting properties of sub-shot-noise spatial intensity
correlations \cite{Jedrkiewicz04,Bondani07,Brida09,Agafonov10} and
macroscopic entanglement
\cite{Munro93,Simon03,Gatti03,Brambilla04,Iskhakov09,Vitelli10}.
More recently, the spectral features of macroscopic twin-beam
states have been investigated in the collinear interaction
geometry close to frequency degeneracy \cite{Spasibko12}.
Moreover, in the high-gain regime, the X-shaped coherence of the
PDC output field \cite{Jedrkiewicz06} and the X-shaped
spatio-temporal twin beam near-field correlations
\cite{Gatti09,Jedrkiewicz12a,Jedrkiewicz12b}, originating from the
space-time coupling in the phase-matching, have been demonstrated.
The high-gain PDC process is also in the focus of attention for
its potential applications. For instance, high-gain PDC has been
used for quantum imaging \cite{Brida10}, ghost imaging
\cite{Bondani12}, and absolute calibration of photo-detectors
\cite{Brida10b,Agafonov11}. The possibility to involve such bright
states in interactions with material quantum objects (atoms,
molecules, quantum dots) has been also addressed. Also the
application of twin beams to quantum memories has been recently
suggested \cite{Gerasimov12}.
\newline
\indent In this paper, we present a joint investigation of spatial
and spectral features of twin-beam states produced in the
high-gain regime with non-negligible pump depletion. The coherence
properties at different values of pump mean power are inferred by
evaluating the intensity auto- and cross-correlation functions on
single-shot images of the far-field ($\theta,\lambda$)
speckle-like pattern of the PDC radiation. 
The initial increase with PDC gain of
the size of coherence areas, both in spectrum and
space, gradually stops and, at a certain pump power, is replaced
by a decrease. This behavior is due to the occurrence of a
progressive pump depletion, which is testified by the evolution of
the spatial and spectral pump beam profiles. A possible explanation can be
given in terms of the varying population of Schmidt paired
modes \cite{Law04} that `diagonalize' the nonlinear interaction.
The number of effectively populated modes is experimentally
accessed by the measurement of spatio-spectral intensity $ g^{(2)}
$-correlation function.
\begin{figure}[h!]
\includegraphics[width=8.5cm]{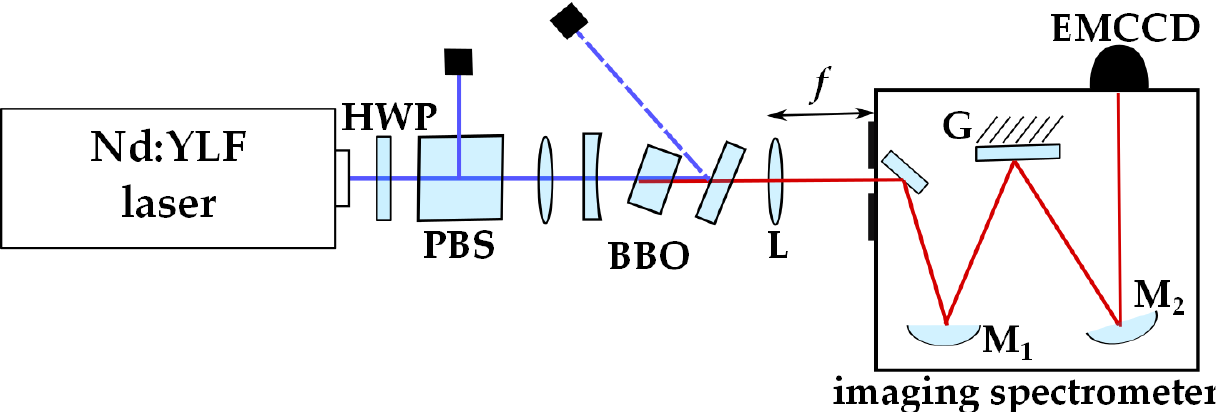}
\caption{(Color online) Experimental setup used for the
spatio-spectral measurements of the twin beam. HWP: half-wave
plate; PBS: polarizing cube beam splitter; BBO: nonlinear crystal;
L: lens, with 60-mm focal length; M$_j$: spherical mirrors; G:
grating; EMCCD: electron-multiplying camera.} \label{fig_setup}
\end{figure}

\begin{figure}[h!]
\includegraphics[width=9cm]{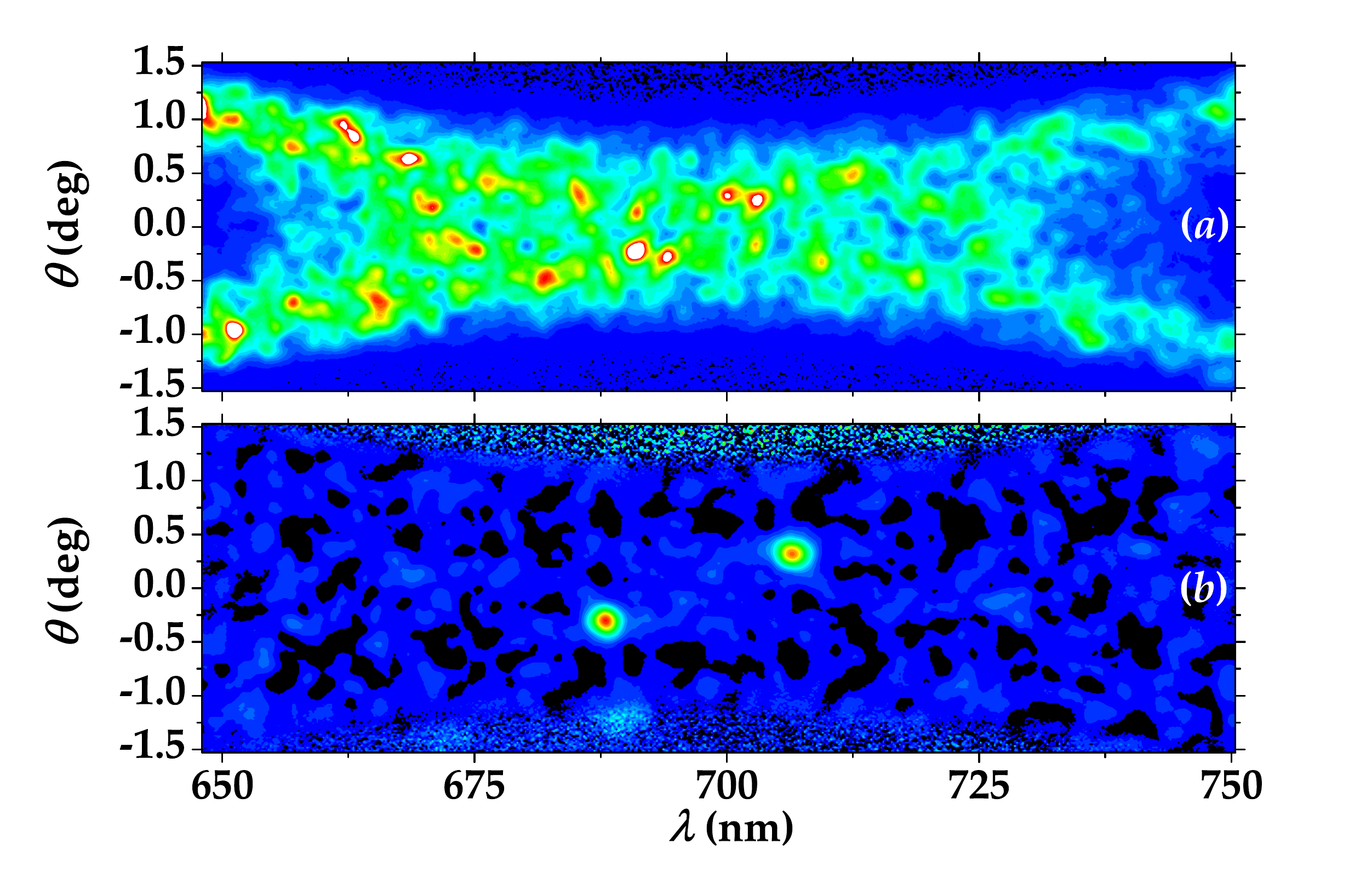}
\caption{(Color online) (a) Single-shot image recorded by the
EMCCD camera, in which the typical speckle-like pattern of PDC in
the spatio-spectral domain is clearly evident. (b) Typical example of
the correlation matrix $\Gamma^{(i,j)}_{k,l}$, in which the auto-
and cross-correlation areas are clearly evident on the left and on
the right side, respectively.} \label{fig_xwaves}
\end{figure}

\indent The experimental setup used for the measurement of the PDC
light structure in the angular and spectral
($\theta,\lambda$) domain is shown in Fig.~1. A type-I 8 mm-long
$\beta$-Barium-Borate (BBO) crystal (cut angle = 37 deg) was
pumped by the third-harmonic pulses (349 nm, 4.5-ps pulse
duration) of a mode-locked Nd:YLF laser (High-Q-Laser),
regeneratively amplified at 500 Hz. The crystal was tuned to have
phase-matching at frequency degeneracy in quasi-collinear
configuration. The pump mean power was changed by means of a
half-wave plate followed by a polarizing cube beam splitter. The
broadband PDC light was collected by a 60-mm focal length lens and
focused on the plane of the vertical slit of an imaging
spectrometer (Lot Oriel) having a 600 lines/mm grating. The
angularly dispersed far-field radiation was then recorded in
single shot by a synchronized EMCCD camera (iXon Ultra 897,
Andor), operated at full frame resolution (512x512 pixels,
16-$\mu$m pixel size).
A typical speckle-like pattern recorded in the quasi-collinear
phase-matching configuration is shown in Fig.~\ref{fig_xwaves}(a).
The existence of intensity correlations between the signal and
idler portions of the twin beam is well-supported by the presence
of symmetrical speckles around the degenerate wavelength and the
collinear direction.\\
\begin{figure}[h!]
\includegraphics[width=9cm]{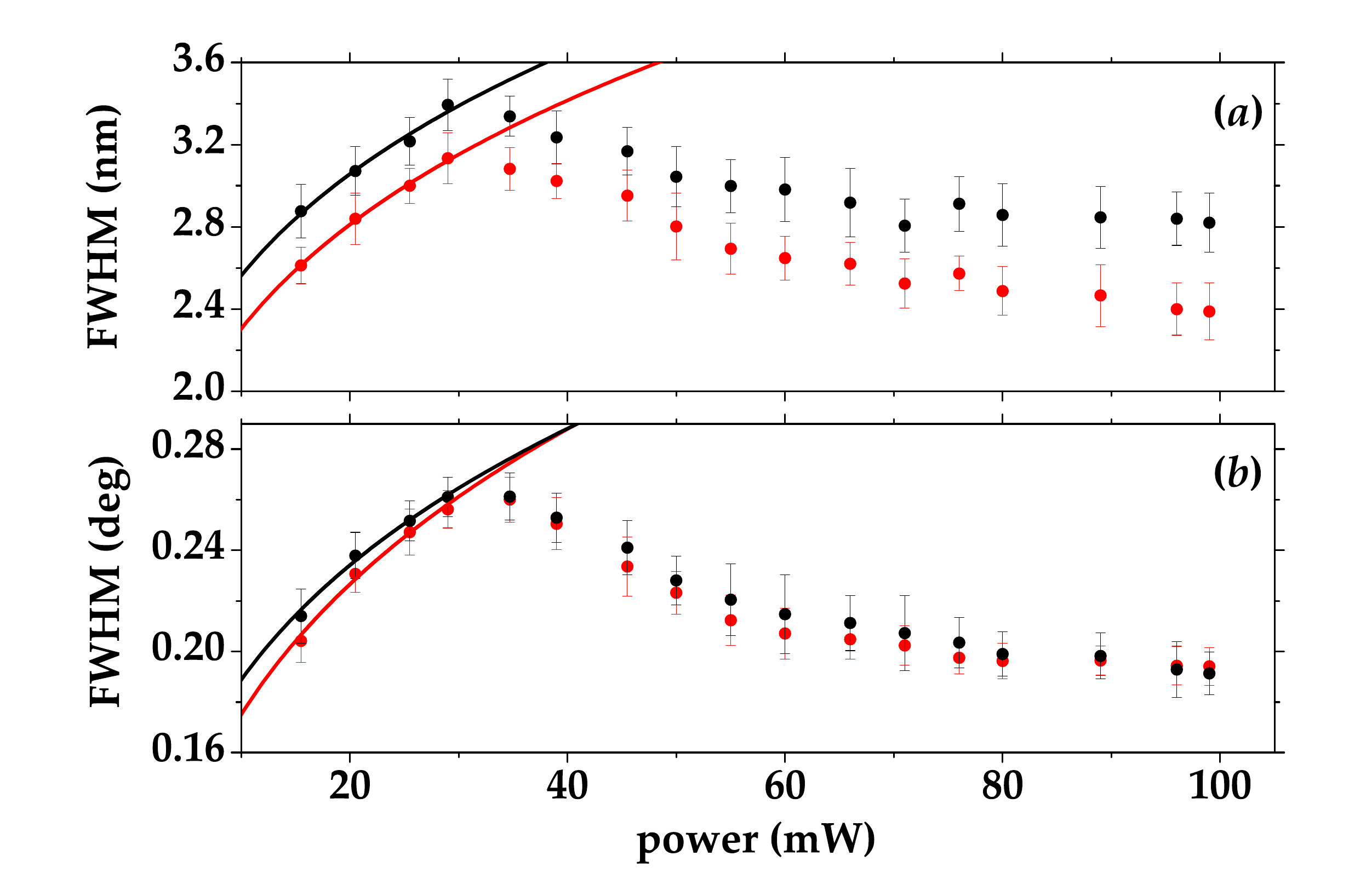}
\caption{(Color online) Evolutions of the spectral (a) and spatial
(b) FWHM size of the second-order auto-correlation (red circles)
and cross-correlation (black circles) functions measured from the
($\theta,\lambda$) spectra of the twin beam as functions of the
pump mean power. A fourth-square root function, as expected from
the theory of PDC structure under the assumption of un-depleted
pump beam, is used to fit the first part of each data set.}
\label{fig_2}
\end{figure}
\begin{figure}[h!]
\includegraphics[width=9cm]{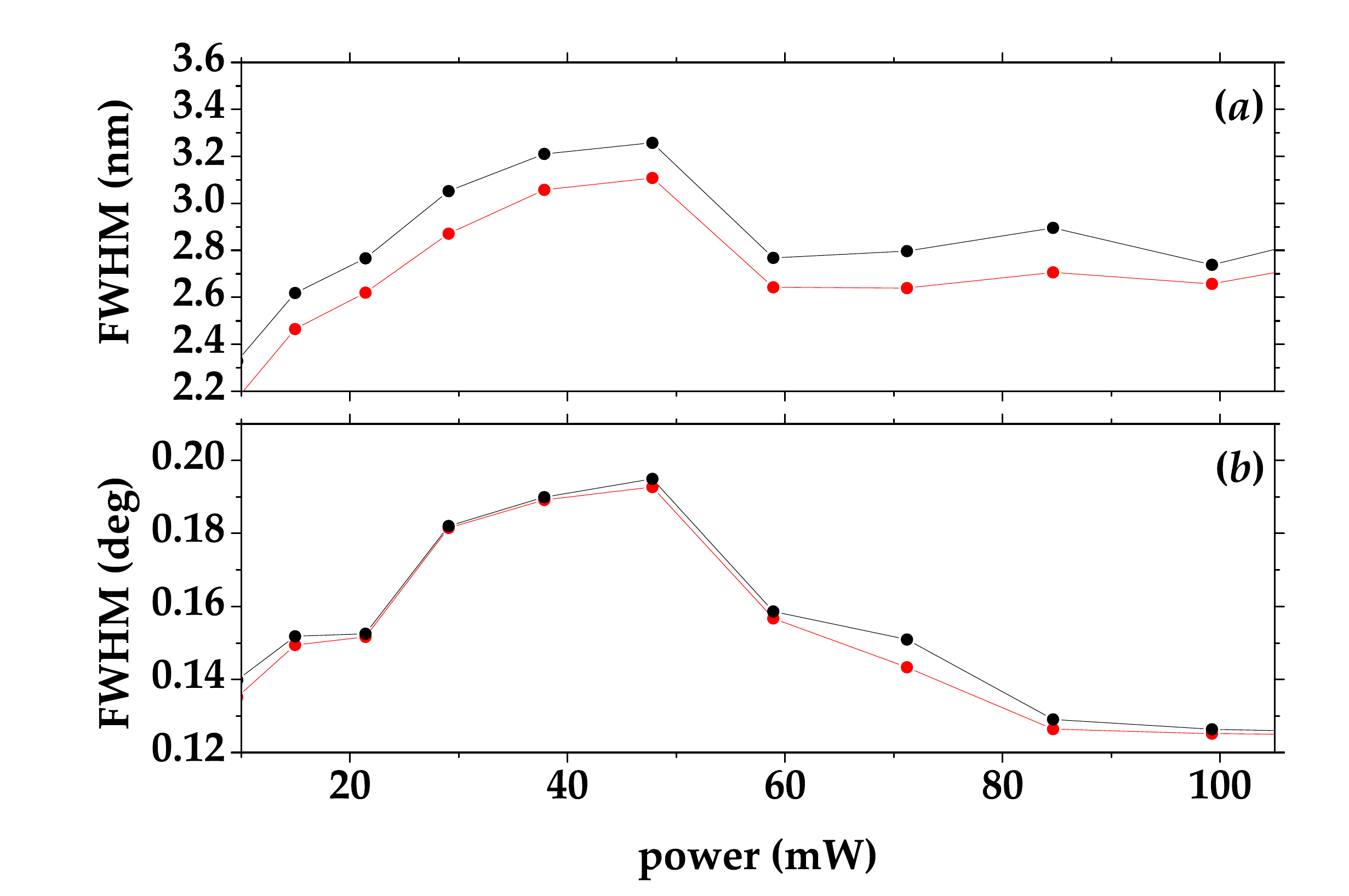}
\caption{(Color online) Simulations of the evolution of the
spectral (a) and spatial (b) FWHM size of the second-order
auto-correlation (red circles+line) and cross-correlation (black
circles+line) functions of the twin beam as functions of the pump
mean power. See the text for details.} \label{fig_simul}
\end{figure}
The evolution of the patterns at different pump mean powers, and
hence at different PDC gains, was investigated by calculating the
spatio-spectral intensity correlation function between a single
pixel at coordinates $(i,j)$ and all the pixels $(k,l)$ contained
in a single image
\begin{equation} \label{correlation}
\Gamma^{(i,j)}_{k,l}= \frac{\langle I_{i,j} I_{k,l} \rangle}{\langle I_{i,j} \rangle \langle I_{k,l} \rangle},
\end{equation}
where $I$ is the intensity value of each pixel expressed in
digital numbers and upon subtraction of the mean value of the
noise measured with the camera in perfect dark, whereas $\langle ...
\rangle$ indicates the averaging over a sequence of 1000
subsequent images. The procedure was applied to a set of pixels
having the abscissa $i$ close to frequency degeneracy and the
ordinate $j$ in the quasi-collinear direction. The function
$\Gamma^{(i,j)}_{k,l}$ defined in Eq.~(\ref{correlation}) is a
matrix having the same size as the original images and containing
both the auto- and the cross-correlation areas (see
Fig.~\ref{fig_xwaves}(b)). The horizontal section of these
correlation areas is related to spectrum, whereas the vertical
section gives information about the angular dispersion. In
Fig.~\ref{fig_2} we show the behaviors of the spatial, $i.e.$ in angular domain, (a) and
spectral (b) widths, full width at half maximum (FWHM), of the
intensity auto-correlation and cross-correlation profiles, as 
functions of the input pump mean power. In both panels, we can
observe an initial growth that reaches the maximum at a pump power
of about 30 mW and then decreases. As shown in the figure, only
the first part of the data is well described by a fourth-root
square function of pump power, as predicted by the theory of
coherence areas under the assumption of un-depleted pump beam
\cite{Brambilla04,Bondani12,Brida09}. The second part of our
experimental results (including the peak and the decrease in the
FWHM) clearly indicates that the assumption of un-depleted pump
beam does not hold anymore. In such a situation, also the pump
beam evolves nontrivially and the corresponding equations of
motion for three-mode interaction can be solved only numerically.
In fact, the behavior shown in Fig.~\ref{fig_2} is qualitatively
reproduced by the numerical simulations presented in
Fig.~\ref{fig_simul}. These results have been obtained by means of
a full 3D+1 (three spatial dimensions + time) numerical modelling
of the optical system based on a pseudospectral (split-step)
integration method of the nonlinear propagation equations
\cite{Brambilla04}. We notice that the slight discrepancy between
the absolute values of the experimental FWHMs and those obtained
from simulations is mainly due to the uncertainty in the
correct positioning of the EMCCD camera in the imaging exit plane of the spectrometer.\\
\indent
The experimental confirmation of the occurrence of pump depletion is given by the evolution
of the spectral and spatial pump-beam profiles shown in the panels of Fig.~\ref{fig_pump}.
\begin{figure}[ht!]
\centering
\includegraphics[width=9cm]{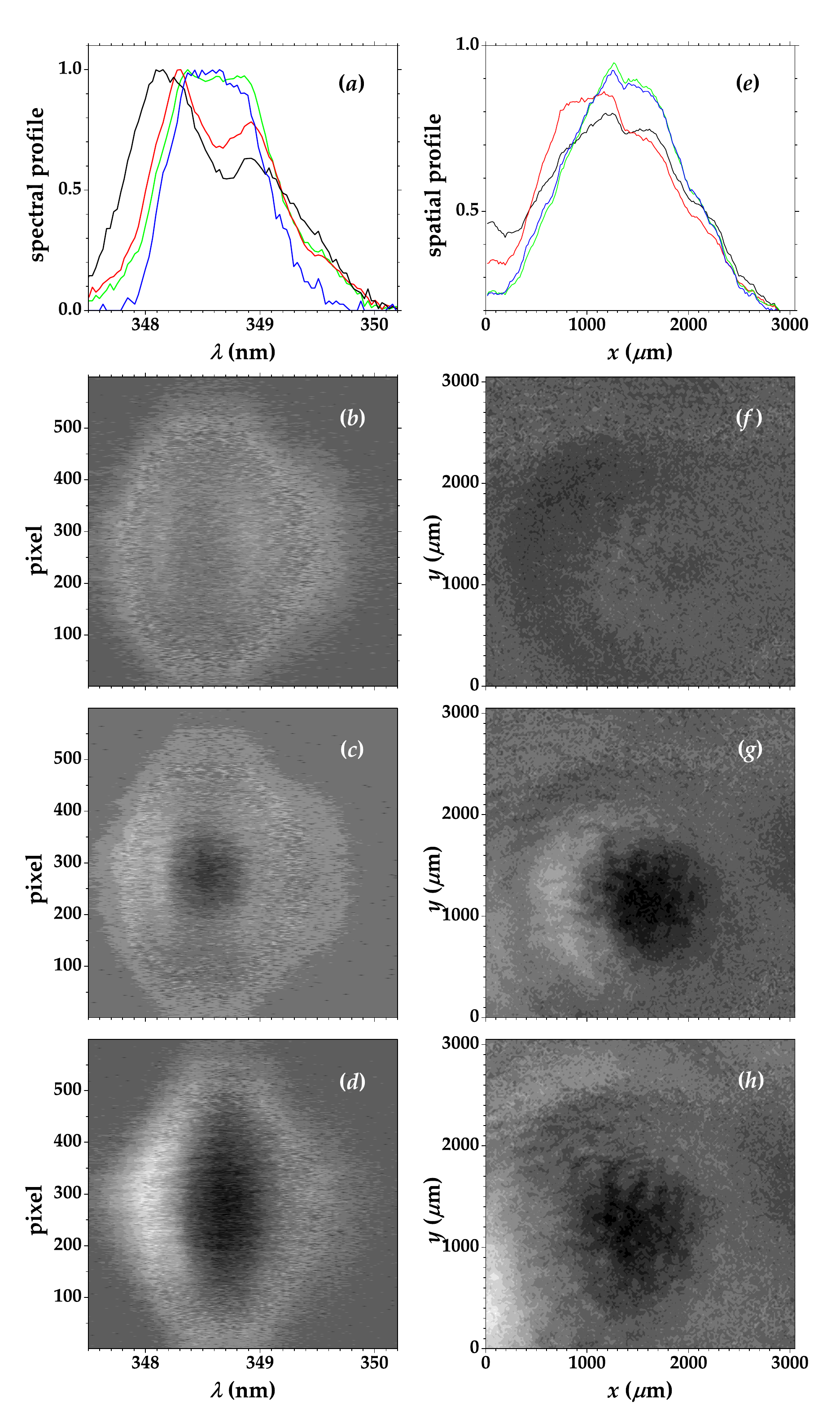}
\caption{(Color online) Spectral and spatial pump beam profiles
for different values of the pump mean power. ($a$): sections of
the spectral profile at different pump powers (blue: 15~mW, green:
35~mW, red 55~mW and black 99~mW), ($b$)-($d$): maps of the
spectral distributions upon subtraction of the distribution of
the least intense measurement (($b$) 35~mW, ($c$) 55~mW and ($d$)
99~mW), ($e$): sections of the spatial profile at different pump
powers (blue: 15~mW, green: 35~mW, red 55~mW and black 99~mW),
($f$)-($h$): maps of the spatial distributions upon subtraction
of the distribution of the least intense measurement.}
\label{fig_pump}
\end{figure}
We obtained a spectral profile of the pump by producing a
magnified image of the near field of the pump on the slit of the
spectrometer and using a CCD camera (DCU223M, Thorlabs, 1024x768
pixels, 4.65-$\mu$m pixel size) to collect the light at the
output. Figure~\ref{fig_pump}(a) displays different sections,
normalized at their peaks, corresponding to different pump mean
power values. First of all, we observe that the spectrum of the
pump turns out to be $\sim$1 nm wide, despite
the longer pulse duration of the non-transform-limited pump
beam. Secondly, we note that both the dips in the sections of
panel (a) and the progressive appearance of a central hole in the
contour plots shown in panels (b)-(d) are a clear signature of
pump depletion. The sections of the spatial profiles presented in
Fig.~\ref{fig_pump}(e) and normalized at their area were obtained
by taking 1:1 images of the pump beam at the output of the crystal
with the same DCU223M camera at different values of the
power. Also in this case a clear dip occurs. It becomes broader and
deeper as the pump mean power increases. Its generation is
initially slightly lateral with respect to the center because of
the pump beam walk-off inside the crystal [see panels (f)-(h)].\\
\indent
The depletion of the pump is also responsible for the evolution of
the total number of photons generated in each realization of the
PDC process. In Fig.~\ref{fig_4} we plot the mean number of
photons detected in an area close to frequency degeneracy and in
the quasi-collinear interaction geometry as a function of the
square root of the pump peak power per pulse. To obtain this
result, we have taken into account the calibration of the camera
sensitivity (5.4 electrons per digital number), its detection
efficiency ($\sim 90\%$ at 698 nm) and all the optical losses.
The mean number of photons shown in Fig.~\ref{fig_4} starts increasing exponentially,
as expected for high-gain PDC under the hypothesis of un-depleted
pump beam. This initial behavior is emphasized in the inset of the same
picture, where the experimental data corresponding to the lowest
pump power values are presented together with the fitting curve
function, $y=A \sinh^2(B x)$. The fitted values of A and B have
been used to calculate the red curve shown in the main figure,
which represents the expected gain behavior in the absence of pump
depletion. In this condition, the gain of the process would vary
from 5.3 up to 13.4, but the occurrence of a progressive depletion
process prevents the exponential growth.
\begin{figure}[h!]
\includegraphics[width=9cm]{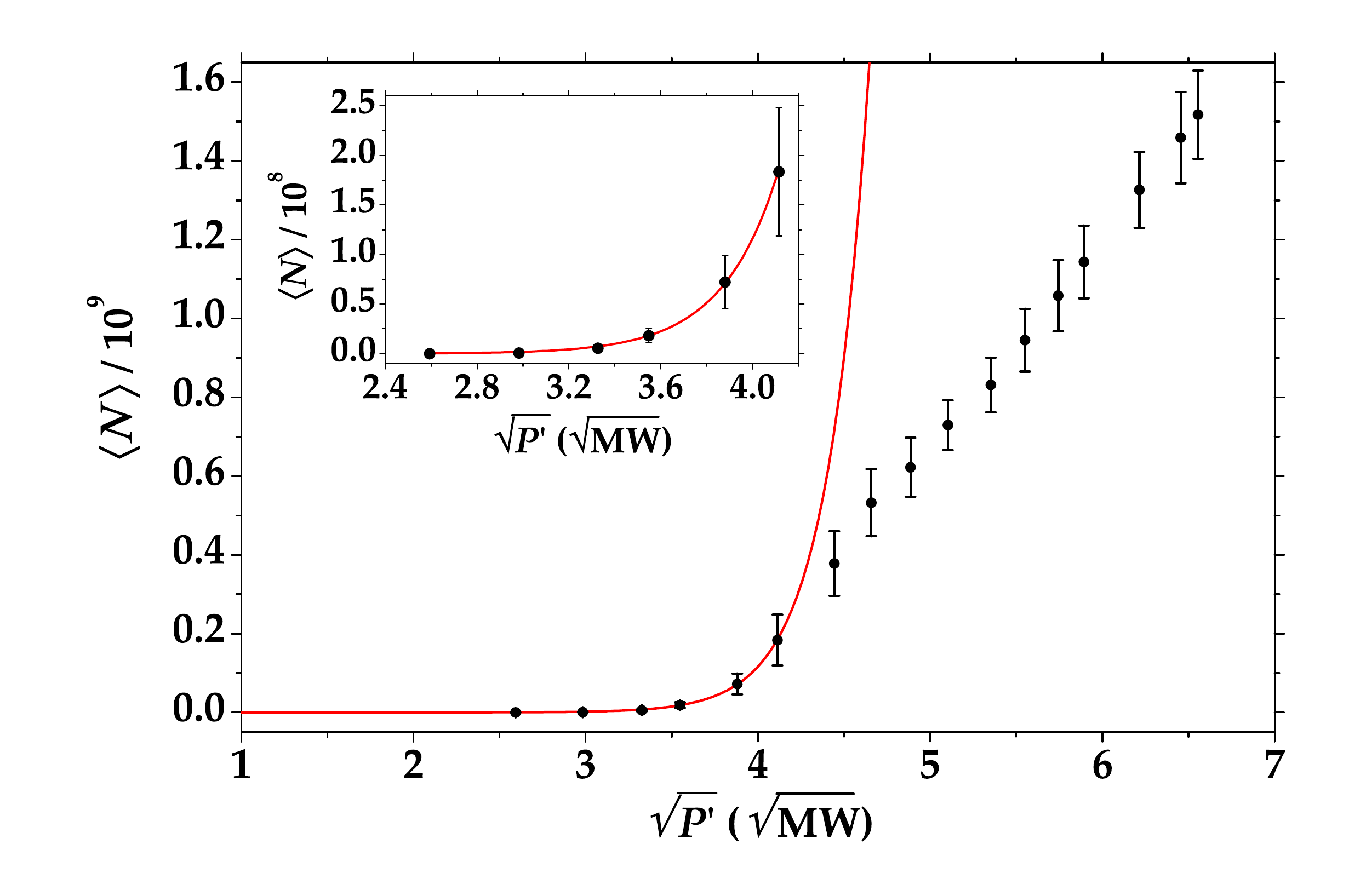}
\caption{(Color online) Evolution of the number of photons
generated by PDC as a function of the square root of the pump peak power $P'$ per pulse.
The theoretical curve (red line) holding under the assumption of
an un-depleted pump beam is also shown. Inset: First part of the
data shown in the main figure (black circles) together with the
fitting curve predicted by the theory (red line).} \label{fig_4}
\end{figure}
\indent
The qualitative change in the evolution of the mean number of
photons, as well as of the size of coherence areas, due to pump
depletion can be described in terms of the modes of radiation
field. When the PDC process occurs at gain values leading to
depletion, also the pump beam evolves in the nonlinear interaction
and the dynamics of the system becomes more complex. In
particular, there is a dependence of the number of effectively
populated signal and idler radiation modes on the pump power. As
the pump power increases, the PDC gain profile becomes narrower
and narrower, and thus signal and idler fields are dominantly
emitted into a smaller and smaller number of modes that gain
energy to the detriment of the others \cite{Wasilewski06,Perez14}.
For sufficiently high values of the pump power, the process of
mode selection reverts as the pump profile undergoes
depletion. For this reason, the gain of the high-populated
low-order modes is on the one side reduced, whereas the gain of
low-populated higher-order modes is on the other side supported.
Such a behavior explains the narrowing of the spatio-spectral
correlation areas shown in Fig.~\ref{fig_2}. The description in
terms of populated radiation modes also explains the slight
discrepancy between auto- and cross-correlation intensity
functions plotted in Fig.~\ref{fig_2}. In fact, cross-correlation
function reflects the mutual coherence between signal and idler
and originates in the pairwise PDC emission, whereas
auto-correlation function expresses the internal coherence due to the
presence of three evolving fields. As such, it is more sensitive
to losses in the modes selection
\cite{speckle_manu}.\\
\indent
To give a quantitative evaluation of such modes, we consider the
well-established theory of PDC at single-photon level, according
to which a bipartite state of biphotons can be written as a sum of
factorized terms \cite{Law04,Christ11} $|\Psi_{12} \rangle =
\sum_k \sqrt{\lambda_k} |u_k\rangle |v_k\rangle$. Here,
$|u_k\rangle$ and $|v_k\rangle$ represent the eigenvectors of a
complete and orthonormal dual basis (the so-called Schmidt modes),
and $\lambda_k$ are the corresponding eigenvalues. We note that
the number of significant eigenvalues, defined as the Schmidt
number or cooperativity parameter, $K=1/ \sum_k \lambda^2_k$,
quantifies the entanglement of the system. For more intense
fields, the definition of cooperativity parameter $ K $ is still
valid provided that we replace parameters $ \lambda_k $ by actual
occupation probabilities of photons in mode $ k $
\cite{Gatti12,PerinaJr2013}. Using this definition, the parameter $
K $ can easily be determined from the normally-ordered
second-order intensity auto-correlation function $ g^{(2)} $
\cite{Perina1991}, along the formula
\begin{equation} \label{cooperativity} g^{(2)} =
1+ 1/K.
\end{equation}
We remark that in the macroscopic regime $g^{(2)}$ has the same expression 
for detected photons, thus allowing the experimental access to the correlation 
function \cite{Allevi12}.
In Fig.~\ref{fig_schmidt}(a) we present the dependence of the
maximum values of spatio-spectral auto- and cross-correlation
functions on increasing values of the pump mean power. We note
that the functions are evaluated both in the signal (colored full
circles) and idler arms (colored empty circles). As already
observed in Fig.~\ref{fig_2} for the size of the auto- and cross-
correlation areas, the plots exhibit a peak for the pump power of
$30$ mW, that lies at the beginning of the pump depletion regime.
Using Eq.~(\ref{cooperativity}) for auto-correlation function $
g^{(2)} $, we estimate the number $ K $ of modes, as shown in
Fig.~\ref{fig_schmidt}(b). The comparison of Figs.~\ref{fig_2} and
\ref{fig_schmidt}(b) confirms the complementary behavior of sizes
of spatio-spectral areas and numbers of modes. Also a close
similarity between the number $ K $ of modes assigned to the
signal and to the idler arms is noticeable. At low pump powers,
where no filters were used to attenuate the light of twin beams,
the numbers $ K $ are nearly identical.
\begin{figure}[h!]
\includegraphics[width=9cm]{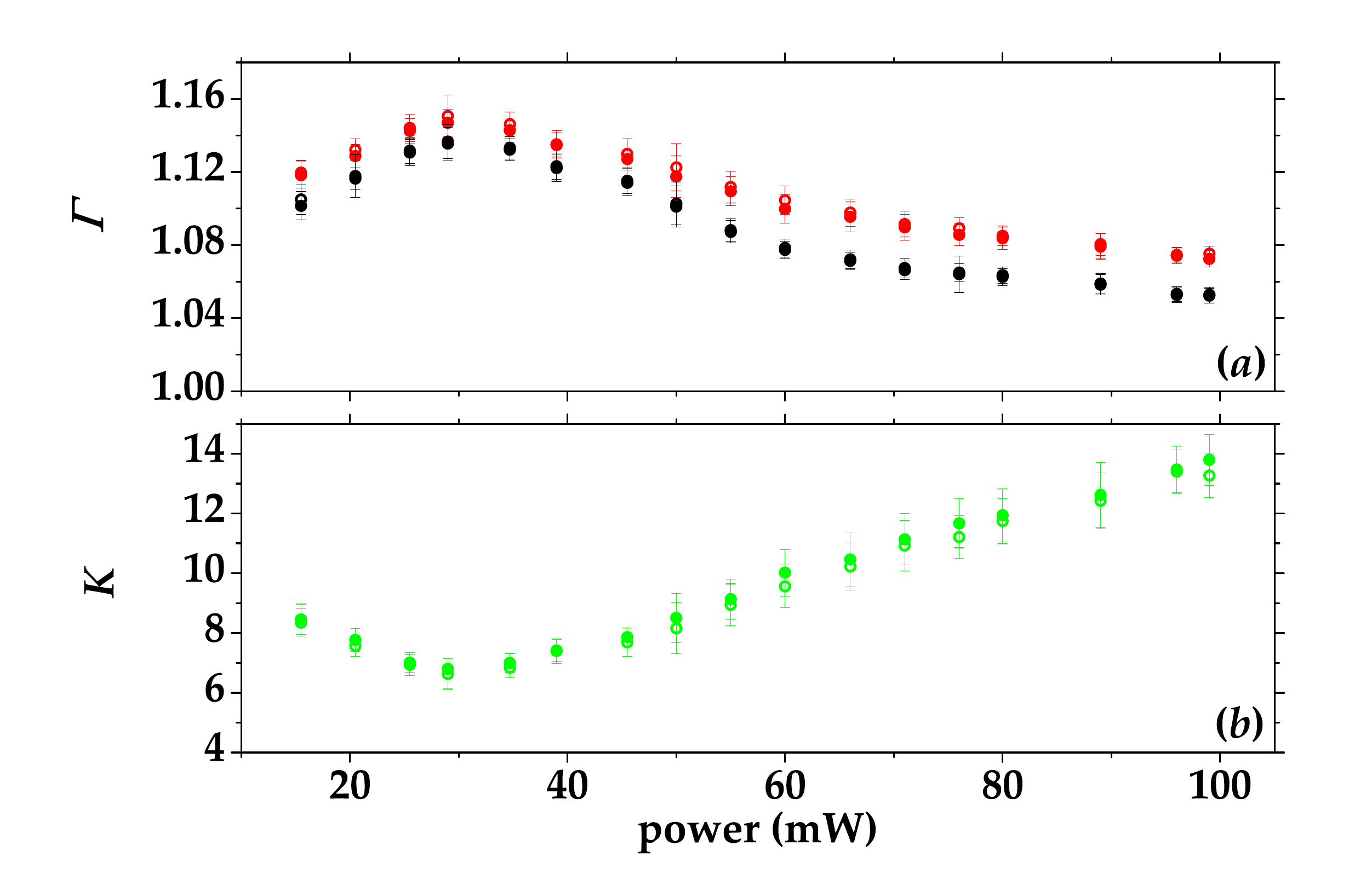}
\caption{(Color online) (a): peak intensity of the spatio-spectral
auto-correlation (red circles) and cross-correlation (black
circles) functions as a function of the pump mean power. (b):
number $ K $ of spatio-spectral modes determined from $ g^{(2)} $-
intensity correlation function. In both panels the data shown as
full (empty) circles characterize the signal (idler) arm.}
\label{fig_schmidt}
\end{figure}
We note that the dual basis revealed in the Schmidt decomposition
\cite{Straupe11,Avella14} can successfully be replaced for more
intense fields by the input-output eigenmodes of the Bloch-Messiah
reduction of the signal-idler unitary evolution operators
\cite{PerinaJr2013}.\\
\indent In conclusion, we have presented an experimental
investigation of the spatio-spectral properties of PDC in the
high-gain regime including pump depletion. The evolution of the
pump is responsible for the qualitative change in the size of auto-
and cross-correlation areas for increasing pump power.
While the correlation areas
gradually broaden for lower pump power values, they undergo narrowing
at increasing high pump powers. We explain this behavior by
the change in the number of effectively-populated paired modes,
whose complementary behavior compared to the sizes of correlation
areas has been experimentally confirmed. These results provide a
clear evidence that the properties of nonlinear processes at high
intensities reflect a complex internal mode structure, which is on
the contrary well-known for fields with intensities at the
single-photon level. Work is in progress to develop a suitable detailed theoretical model that will
be very useful to reach a deeper insight into the nonlinear process.
\newline
\indent
\\
This work was supported by MIUR under the grant agreement FIRB
ÒLiCHISÓ - RBFR10YQ3H. JP and OH acknowledge projects P205/12/0382
of GA \v{C}R and projects CZ.1.05/2.1.00/03.0058 and
CZ.1.07/2.3.00/20.0058 of M\v{S}MT \v{C}R. OJ thanks Paolo Di Trapani for fruitful discussions.

\end{document}